\documentclass[12pt]{article}
\usepackage{amsfonts,amssymb,amscd}

\newcommand{\be}{\begin{equation}\label}
\newcommand{\ee}{\end{equation}}

\newcommand{\p}{\prime}
\newcommand{\bib}{\bibitem}

\begin{document}

\begin{center} 

{\bf ON THE STRUCTURE OF GENERAL SOLUTION\\
TO THE EQUATIONS OF SHEAR-FREE NULL CONGRUENCES}

\vskip4mm

{\bf Vladimir V. Kassandrov} 

\vskip4mm

{\it (Institute of Gravitation and Cosmology, \\ 
Russian People's Fiendship University, Moscow, Russia)}

\end{center}

\vskip4mm
\noindent
{\bf Abstract.} We demonstrate that the source of a shear-free null congruence is, 
generically, a world sheet of a singular string in the complex Minkowski space. 
In particular 
cases the string degenerates into the Newman's point singularity. We describe 
also a time-dependent deformation of the Kerr congruence which could indicate 
the unstability of the Kerr (Kerr-Newman) (electro)vacuum solution. Then we 
consider a 5D extension of Minkowski space obtained by complexification of the 
time coordinate. Singular locus of a shear-free null congruence on such an extended  
space-time consists of a (great) number of point particles 
(``markeons'') and 2D surfaces -- light-like wave fronts.

\vskip6mm

Congruences of light rays, with zero shear in particular, play crucial role in 
general relativity and in twistor aspects of geometry of flat Minkowski 
space-time $\bf M$. They relate also to the conditions of differentiability 
for the biquaternion-valued functions which may be put into correspondence with 
fundamental physical fields (in the framework of the so called 
{\it algebrodynamical} approach, see below). 

General solution to the defining equations of shear-free (null geodesic) 
congruences (SFC) is represented by the {\it Kerr theorem} and can be 
formulated in the following implicit algebraic form~\cite{1}:
\be{Kerr}
\Pi (\xi, iX\xi) = 0, 
\ee
where $\Pi(\xi,\tau)$ is an {\it arbitrary homogeneous} (and almost everywhere 
analytical) function of three {\it projective} components of the twistor $\{\xi,
\tau\}$ on $\bf M$, i.e. of a pair of the 2-spinors linked at every 
space-time point $X=X^+=\{X_{AA^\p}\},~A=A^\p= 0,1$ via the Penrose 
{\it incidence relation}, 
\be{inc}
\tau=iX\xi~~~(\tau^A=iX^{AA^\p}\xi_{A^\p}) 
\ee
(see, e.g.,~\cite{1}, ch 7). 
At a fixed point $X$, resolving (\ref{Kerr}) with respect to the ratio of  
components of the spinor $\xi_A=\xi_A (X)$, one comes to a null  
4-vector field $k_\mu = \xi_A \xi_{A^\p},~\mu= 0,1,2,3$ tangent to the 
congruence which is necessarily shear-free; according to the 
Kerr theorem, all (analytical) SFC can be constructed via the above 
presented algebraic procedure. 

{\it Caustics} - singularities of SFC - are defined by the  condition 
\be{caust}
\Pi^\p (\xi) = 0,
\ee 
i.e. correspond to the space-time points at which the {\it total derivative} 
of $\Pi$ with respect to the {\it ratio} of spinor components turns to zero. 
Eliminating the latter from  eq.(\ref{caust}) and substituting the 
result into eq.(\ref{Kerr}), one comes to the {\it equation of motion} 
\be{eqmotion}
S(X)\equiv \Pi(\xi(X), iX\xi(X)) = 0  
\ee
which defines the shape of singular locus of the congruence (at a fixed moment 
of time) and its time evolution as well. It has been shown~\cite{3,4} that 
any such function $S(X)$ satisfies the {\it complex eikonal equation} 
(CEE).

In the framework of {\it algebrodynamics} (see~\cite{2,3,4} and references 
therein) such singularities (when bounded in 3-space) were identified with 
{\it particle-like} formations. A remarkable example of these in general 
relativity is represented by the Kerr's {\it singular ring} -- the caustic of 
the twofold static and axisymmetrical Kerr congruence. It possesses some of the quantum 
quantum numbers of elementary fermion (Dirac electron), see,e.g.~\cite{5}. Newman
et al.~\cite{6,7} have shown that the Kerr-like congruences may be viewed as 
being generated by a ``virtual'' point charge moving along a complex world line 
$Z_\mu = Z_\mu (\sigma),~\sigma \in \mathbb{C}$ in the {\it complex extension} 
$\mathbb{C}\bf M$ of Minkowski space-time $\bf M$ and therewith emanating 
rectilinear complex ``light-like rays''. Restriction of such a complex 
congruence to the real $\bf M$ gives there rise to a SFC, to the Kerr one in 
particular (the latter in the case of virtual charge at rest). 

However, this construction~\cite{8} covers only a narrow subclass of SFC. Below 
we present the theorem which specifies the source of a SFC of a {\it generic 
type}. Precisely, we define the source of a SFC as the {\it hyper-caustic} set, 
i.e. the singular locus at which, apart from the conditions (\ref{Kerr}) and 
(\ref{caust}), the {\it second} total derivative (with respect to the ratio of 
spinor components) turns to zero,
\be{cusp}
\Pi^{\p\p}(\xi) = 0,
\ee
so that one deals here with the singularities of caustics -- the {\it cusps}. 
The theorem can be formulated as follows.

\bigskip
\noindent
{\bf Theorem.}
 
\noindent
The source of a shear-free congruence of a generic type is a {\it complex string} 
$Z_\mu = Z_\mu (\lambda,\sigma),~\lambda,\sigma \in \mathbb{C}$,  i. e. a 
1-dimensional complex curve $Z_\mu = Z_\mu (\lambda)$ evolving in ``complex 
time'' $\sigma$. This is determined by the set of algebraic eqs.
(\ref{Kerr}), (\ref{caust}), (\ref{cusp}). Precisely, at any point $Z\in 
\mathbb{C}\bf M$ of complex Minkowski space 
the value of (projective) twistor $\{\xi,\tau\}$ of a SFC is the same as at 
some point of the generating string $Z_\mu = Z_\mu (\lambda, \sigma)$, and 
correspondent point belongs to the complex light cone of the string element. 

\bigskip
\noindent
{\bf The proof} is very simple. Indeed, system of the {\it three} 
eqs.(\ref{Kerr}), (\ref{caust}), (\ref{cusp}) together with the {\it two} 
incidence constraints (\ref{inc}), in all of which in the considered case of 
full complex space $\mathbb{C}\bf M$ the substituition of coordinates 
$iX \mapsto Z$ is made, determine, generically, {\it five} of the {\it seven} 
unknowns -- {\it four} complex coordinates $Z^{AB}$ and {\it three} projective 
twistor components. In result, as a solution of the joint system one obtains   
\be{stringtau}
Z_\mu = Z_\mu (\tau^1,\tau^2),
\ee
where some two twistor components, say $\tau^A$, are taken to parametrize the 
world sheet of the complex string whereas the third one, say $G=\xi_2 / \xi_1$, 
is then defined by the basic Kerr constraint (\ref{Kerr}). Of course, 
generally any other parametrization of the world sheet could be chosen. 

Let now a ``branch'' of the (generally multi-valued) projective twistor field 
$\{\xi,\tau\}$ of a SFC is obtained at a point $Z\in \mathbb{C}\bf M$ as a 
solution of the basic Kerr eq. (\ref{Kerr}). Then eq. (\ref{stringtau}) immegiately 
specifies an element of generating string at which all of the components of 
the twistor field take the same values as at the considered point $Z$. $\Box$

\bigskip

In other words, if a {\it world-sheet} $Z_\mu (\lambda,\sigma)$ of 
complex string is known, the congruence is easily reconstructed both on 
$\mathbb{C}\bf M$ and $\bf M$ via examination of rectilinear complex 
null ``rays'' emanating by all its elements, and any SFC can be obtained in 
this way. In particular cases (e.g., for axisymmetrical congruences) the 
source string degenerates into the Newman's point charge. Moreover, for some 
types of SFC the world  sheet of generating string can {\it coexist} 
with the world line of the Newman's point charge -- {\it focal line} 
of the congruence.  Examples of such congruences, perhaps the most 
interesting ones, will be presented elsewhere. 

Simple considerations on codimensions show that, generically, the {\it 
caustic} locus on $\bf M$ consists of a number of {\it real isolated 
strings}, usually non-bounded and unstable in time. Axisymmetrical Kerr's  
SFC with the Kerr's singular ring is only an exceptional example of a bounded and 
``nonradiating'' caustic. Indeed, by continious change of parameters of the 
Kerr's generating function $\Pi$ it can be deformed to the form 
\be{deform}
\Pi= \tau^2 (1+ih) -G\tau^1(1-ih)+2iaG, 
\ee
with a new parameter $h\in \mathbb{R}$ and the parameter $a\in \mathbb{R}$ 
being the old Kerr's one. Then one obtains a time-dependent axisymmetrical 
congruence generated by the Newman's charge moving uniformly in $\mathbb{C}\bf 
M$ with ``imaginary velocity'' $v=ih$. Caustic locus of correspondent SFC in 
$\bf M$ is now represented by an uniformly collapsing and then expanding ring
\be{defring}
z=0, ~~~\rho = \sqrt{x^2+y^2} = v(t-t_0),
\ee
where $t_0 = a/\sqrt{1+h^2}$ and the speed $v=h/\sqrt{1+h^2}$ of 
contraction (at $t<t_0$) or expansion (at $t>t_0$) is always less than the speed 
of light $c=1$. In account of existence of such a deformation of the Kerr 
congruence one can even suspect that the very {\it Kerr (Kerr-Newman) solution 
to Einstein (Einstein-Maxwell) (electro)vacuum system is unstable} as its 
generating SFC does.  

Consider now the {\it hyper-caustic} locus of a generic SFC on $\bf M$. It is  
easy to see, in account of similar considerations on codimension, that it will 
be represented by a system of isolated points -- {\it instantons}
~\footnote{This term here has, certainly, nothing to do with the familiar 
Yang-Millls instantons} -- which appear only at some particular instants  
and dissappear immegiately. 

All this seems to be far from the observable structure of physical matter. 
There is, however, a successive hypothesis  which drastically changes 
the situation. Namely, let us consider a {\it five-dimensional} submanifold 
of $\mathbb{C}\bf M$ formed via restriction onto $\bf M$ of three 
space-like coordinates, $Z_a \to X_a \in \mathbb{R},~a=1,2,3$, whereas 
the time-like coordinate $Z_0 = t + i s,~~t, s\in \mathbb{R}$ is assumed   
to remain essentially complex-valued. On such a 5D cut of $\mathbb{C}\bf M$
singular locus is represented by the 2D surfaces which evolve in {\it real} 
time $t$, whereas the structure of hyper-caustics -- by a (great) number of 
point charges -- {\it markeons} (from English ``mark'') -- moving along 
their world lines, effectively interacting each other and stable by 
definition. As to the singular 2D surfaces, they strongly resemble the 
{\it light-like wave fronts}. 

Thus, we see that assumption of the 5D structure of physical space-time 
(formed by complexification of the time-like parameter) leads to a trustful  
structure of elementary matter with the point-like stable ``markeons'' and 
the singular wave fronts as the basic constituents. Moreover, it can be easily 
demonstrated that the complex null cone (CNC) on such a 5D subspace 
{\it degenerates into the familiar light cone of the 4D Minkowski space-time}. 
Indeed, the CNC equation 
\be{CNC}
(Z_0)^2-(Z_1)^2-(Z_2)^2-(Z_3)^2=0
\ee
for the case of three real and one complex time-like coordinate corresponds to 
two {\it real} constraints from the second of which $2ts=0$ one gets $s=0$ so 
that the first constraint reduces to the real Minkowski light cone 
\be{Mink}
t^2 - x^2 - y^2 - z^2 = 0.
\ee

On the other hand, this means that the fifth 
coordinate $s$ {\it remains constant along any ray of the congruence} 
and does not contribute, therefore, into the observable metrics of physical 
space-time.  In other words, reduction of the extended space to the 4D physical 
one takes here place {\it dynamically}, i.e. despite any known artificial 
assumptions like cylindricity condition, compactification etc. Dynamics of 
markeons and their interaction with singular wave fronts are under study.

\end{document}